\documentclass[a4paper,twoside,10pt]{article}
\input{style.sty}
\begin{document}
%
%%%%%%%%%%
% Headers and page style
%%%%%%%%%%
\pagestyle{fancy}
\fancyhead{}
  \fancyhead[RO,LE]{\thepage}
  \fancyhead[LO]{T. Suzuki}          %% Your name
  \fancyhead[RE]{Estimation of $\xi$ parameter on the Moffat Gravity}
\rfoot{}
\cfoot{}
\lfoot{}
%%%%%%%%%%%%%%%%%%%%
% Page label
%%%%%%%%%%%%%%%%%%%%
\label{P31}                             %% Your presentation number
%%%%%%%%%%%%%%%%%%%%
% Title
%%%%%%%%%%%%%%%%%%%%
\title{%
Estimation of $\xi$ parameter on the Moffat Gravity
}
%
%%%%%%%%%%%%%%%%%%%%
% Author
%%%%%%%%%%%%%%%%%%%%
\author{%
  Takayuki Suzuki \footnote{Email address: n003wa@yamaguchi-u.ac.jp}$^{(a)}$
}
%
%%%%%%%%%%%%%%%%%%%%
% Address
%%%%%%%%%%%%%%%%%%%%
\address{%
  $^{(a)}$Department of Physics, Yamaguchi University, Yamaguchi city, 
       Yamaguchi prefecture 753-8512\\}
%
%%%%%%%%%%%%%%%%%%%%%%%%%%%%%%%%%%%%%%%%%
\abstract{
%%%%%%%%%%%%%%%%%%%%%%%%%%%%%%%%%%%%%%%%%
Scalar Tensor Vector Gravity(STVG) is one of modified gravity theories developed by John Moffat(2005).
MOG is abbreviated name for this theory.It can explain a galactic rotation curve and the structure formation  without dark matter.
It can also explain acceleration universe without dark energy.But,they obtaion only a spherically symmetric, static vacuum solution about MOG.
On this theory,the gravitational field produced by two point sources is not simply the sum of their respective spherically symmetric static vacuum solutions.
However,in arXiv:0805.4774, the method to adapt MOG to extended distribution of matter is described by phenomenalism.
A new parameter $\xi$ is introduced in this phenomenalical description.This paper shows estimation of MOG's $\xi$ parameter.
In coclusion,$\xi$ should be less than $\mathcal{O}(10^2)$ to reproduce ``flat'' rotation curves observed.\\
%%%%%%%%%%%%%%%%%%%%%%%%%%%%%%%%%%%%%%%%%
}
%%%%%%%%%%%%%%%%%%%%%%%%%%%%%%%%%%%%%%%%%

%%%%%%%%%%%%%%%%%%%%%%%%%%%%%%%%%%%%%%%%%
\section{Alternative to ``Dark side of universe'' -Introduction}
%%%%%%%%%%%%%%%%%%%%%%%%%%%%%%%%%%%%%%%%%
The standard model of cosmology today, the $\Lambda$CDM model, provides an excellent fit to cosmological observations.
But,most of the composition  of the universe {\em is invisible and undetectable which is named dark energy or dark matter}.
This fact provides a strong incentive to seek alternative explanations that can explain cosmological observations without dark matter or dark energy.
Modified Gravity (MOG) \citep{Moffat} has been used successfully to account for galaxy cluster masses \citep{Brownstein3}, the rotation curves of galaxies \citep{Brownstein}\citep{Brownstein2}, velocity dispersions of satellite galaxies, and globular clusters \citep{Toth3}. 
It can explain the observation of the Bullet Cluster \citep{Brownstein2} without cold dark matter.
Besides, MOG also meets the challenge posed by cosmological observations. 
In the paper(arXiv:0710.0364)\citep{MofTot2007},it is demonstrated that MOG produces an acoustic power spectrum, a galaxy matter power spectrum, and a luminosity-distance relationship that are in good agreement with observations.

%%%%%%%%%%%%%%%%%%%%%%%%%%%%%%%%%%%%%%%%%
\section{Theory of MOG - Precedent study}
%%%%%%%%%%%%%%%%%%%%%%%%%%%%%%%%%%%%%%%%%
\subsection{Action}
The action of Moffat theory is constructed as follows \cite{Moffat}. 
STVG is formulated using the action principle. In the following discussion, a metric signature of $[+,-,-,-]$ will be used; the speed of light is set to $c=1$, and we are using the following definition for the Ricci tensor:
$R_{\mu\nu}=\partial_\alpha\Gamma^\alpha_{\mu\nu}-\partial_\nu\Gamma^\alpha_{\mu\alpha}+\Gamma^\alpha_{\mu\nu}\Gamma^\beta_{\alpha\beta}-\Gamma^\alpha_{\mu\beta}\Gamma^\beta_{\alpha\nu}.$
We denotes the Einstein-Hilbert Lagrangian:${\mathcal L}_G=-\frac{1}{16\pi G}\left(R+2\Lambda\right)\sqrt{-g},$
where $R$ is the trace of the Ricci tensor, $G$ is the gravitational constant, $g$ is the determinant of the metric tensor $g_{\mu\nu}$, while $\Lambda$ is the cosmological constant.
Introducing the Proca action Maxwell-Proca Lagrangian for the STVG vector field $\phi_\mu$:
\begin{eqnarray}
{\mathcal L}_\phi=-\frac{1}{4\pi}\omega\left[\frac{1}{4}B^{\mu\nu}B_{\mu\nu}-\frac{1}{2}\mu^2\phi_\mu\phi^\mu+V_\phi(\phi)\right]\sqrt{-g},
\end{eqnarray}
where $B_{\mu\nu}=\partial_\mu\phi_\nu-\partial_\nu\phi_\mu$, $\mu$ is the mass of the vector field, $\omega$ determines the strength of the coupling between the fifth force and matter, and $V_\phi$ is a self-interaction potential.
The three constants of the theory, $G$, $\mu$ and $\omega$, are promoted to scalar fields by introducing associated kinetic and potential terms in the Lagrangian density:
\begin{eqnarray}
{\mathcal L}_S=-\frac{1}{G}\left[\frac{1}{2}g^{\mu\nu}\left(\frac{\nabla_\mu G\nabla_\nu G}{G^2}+\frac{\nabla_\mu\mu\nabla_\nu\mu}{\mu^2}-\nabla_\mu\omega\nabla_\nu\omega\right)+\frac{V_G(G)}{G^2}+\frac{V_\mu(\mu)}{\mu^2}+V_\omega(\omega)\right]\sqrt{-g},
\end{eqnarray}
where $\nabla_\mu$ denotes covariant differentiation with respect to the metric $g_{\mu\nu}$, while $V_G$, $V_\mu$, and $V_\omega$ are the self-interaction potentials associated with the scalar fields.
The STVG action integral takes the form $S=\int{({\mathcal L}_G+{\mathcal L}_\phi+{\mathcal L}_S+{\mathcal L}_M)}~d^4x,$
where ${\mathcal L}_M$ is the ordinary matter Lagrangian density.

\subsection{Weak field approximation of MOG}
The field equations of STVG can be developed from the action integral using the variational principle.
First a test particle Lagrangian is postulated in the form${\mathcal L}_\mathrm{TP}=-m+\alpha\omega q_5\phi_\mu u^\mu,$
where $m$ is the test particle mass, $\alpha$ is a factor representing the nonlinearity of the theory, $q_5$ is the test particle's fifth-force charge, and $u^\mu=dx^\mu/ds$ is its four-velocity.
Assuming that the fifth-force charge is proportional to mass,$q_5=\kappa m$, the value of $\kappa=\sqrt{G_N/\omega}$ is determined 
and the following equation of motion is obtained in the spherically symmetric, static and weak gravitational field of a point mass of mass $M$:

\begin{eqnarray}
\ddot{r}=-\frac{G_NM}{r^2}\left[1+\alpha-\alpha(1+r/\lambda)e^{-r/\lambda}\right],\label{PointAcceleration}
\end{eqnarray}
where $G_N$ is Newton's constant of gravitation.
Further study of the field equations allows a determination of $\alpha$ and $\lambda$ for a point gravitational source of mass $M$ in the form:
\begin{equation}
\lambda=\frac{\sqrt{M}}{D},\label{eq:mu}\\
\end{equation}
\begin{equation}
\alpha=\frac{M}{(\sqrt{M}+E)^2}\left(\frac{G_\infty}{G_N}-1\right).\label{eq:alpha}
\end{equation}

The constants $D$ and $E$ various astronomical observation yield the following values:
$D\simeq 6250 M_\odot^{1/2}\mathrm{kpc}^{-1},$ $E\simeq 25000 M_\odot^{1/2},$
In the weak-field approximation, STVG produces a Yukawa-like modification of the gravitational force due to a point source.
Nearby a source gravity,the repulsive force which comes from vector field counteracts attraction.
But far from a source gravity,Yukawa-like repulsive force cannot reach because vector field is massive.
Intuitively, this result can be described as follows: far from a source gravity is stronger than the Newtonian prediction, 
but at shorter distances,gravity is comparable to Newtonian.

\subsection{Adaption MOG to extended distribution of matter}

The gravity acceleration equation of the STVG in a weak gravitational field is shown in the Eq.\ref{PointAcceleration}.
But,This equtaion is adaped only gravitational field of ``a point mass''.
It is not applied simple superposition principle of forces without self-contradiction.
In a theory that offers linear behavior in the weak field limit,for example general relativity ,an extended distribution of matter can be well approximated by a large number of point masses.
However MOG is not so: the gravitational field produced by two point sources is not simply the sum of their respective spherically symmetric static vacuum solutions.
Presently, we do not have solutions to the MOG field equations for extended matter distributions. But, they slove this problem phenomenologically in in Moffat et al.(2009)\cite{Toth5}.
They seek an effective mass function $M_{eff}(\vec{x},\vec{r})$, to be used in place of $M$ in (\ref{eq:alpha}) and (\ref{eq:mu}), that determines an ``effective mass'', to be used in the formulae for $\alpha$ and $\lambda$.
The function should yield the mass of the source in the case of a point source, and a mass proportional to volume in the case of a constant distribution.
One function that satisfies these criteria is in the form,
\begin{equation}
M_{eff}(\vec{x},\vec{r})=\int\rho(\bar{\vec{x}})\exp\left(-\xi\frac{|\bar{\vec{x}}-\vec{x}|}{|\vec{r}-\vec{x}|}\right)~d^3\bar{\vec{x}},\label{eq:76}
\end{equation}
where $\xi$ is a parameter which should be decided to satisfy a astronomical phenomenon observed.

\section{Estimation of $\xi$ parameter by N-body calculation - Method}

We let the Eq.\ref{eq:76} become disintegration for N-body simulation. And the following expressions are derived by adapting itself to many body system.
In the following expression,``i-particle'' is gravity source particle,``j-particle'' is received particle.
\begin{eqnarray}
a_j=-\sum_i^N \frac{G_N m_i (\vec{r_i}-\vec{r_j})}{(\left|r_i-r_j \right|^2+eps^2)^{3/2}}G_{eff}(i,j) \label{TwobodyAcceleration}
\end{eqnarray}
\begin{eqnarray}
G_{eff}(i,j)= \left[1+\alpha_{ij}-\alpha_{ij}(1+\frac{\left| r_i-r_j \right|}{\lambda_{ij}})e^{-\left| r_i-r_j \right|/\lambda_{ij}}\right]\label{Geff_ij}
\end{eqnarray}
\begin{eqnarray}
\lambda_{ij}=\frac{\sqrt{M_{eff}}}{D},\alpha_{ij}=\frac{19M_{eff}}{(\sqrt{M_{eff}}+E)^2},M_{eff}(i,j)=\sum_l^N m_l exp(-\xi \frac{\left| r_l-r_i \right|}{\left| r_i-r_j \right|})
\end{eqnarray}

This calculation needs the information of all the particles whenever we calculate force between two particles. That is why the calculation number of times becomes $\mathcal{O}(N^3)$.
By the normal method, we cannot carry out such an enormous calculations.But,We use ``mass shell and G shell aproximation'' in our precedent study \citep{my paper1}.
Then,we can reduce the calculation number of times to $\mathcal{O}(N^2)$.
To estimate value of $\xi$ which can reproduce observational phenomena ,we calculated the acceleration which each particle of the expnential disk.
The following is parameter of the disk.Thease are comparable to the Milky way.
\begin{table}[htbp]
\begin{center}
\begin{tabular}{ccc}
\hline
1R(radius) & 25kpc \\
1M(mass) & $5\times 10^{10} M_\odot$ \\
1T(time) & 0.26Gry \\
1V(velocity) & 92km/s \\
\hline
\end{tabular}
\caption{The standardization of the unit}
\end{center}
\end{table} 
\begin{table}[htbp]
\begin{center}
\begin{tabular}{ccc}
\hline
$R_d$(scale length of disk) & 2.5kpc \\
$R_{cut}$(radius of disk) & 25.0kpc \\
$Z_d$(scale length of disk height) & 0.5kpc \\
$M_d$(total mass of disk) & $5\times 10^{10} M_\odot$ \\
eps(softening length) & 0.048kpc \\
Nbody(number of particles) & 10000 \\
\hline
\end{tabular}
\caption{Parameters of the disk}
\end{center}
\end{table} 
This $\xi$ is the non-dimensional parameter which occured from phenomenological method.
Theoretically,it is no problem that either $\xi=10^{-10}$ or $\xi=10^{10}$.Then,we can't guess the rough size of $\xi$.
We perform order-estimation about $\xi$. We examined 7 models:$\xi=10^{-2},10^{-1},10^0,10^1,10^2,10^3,10^4$.

\newpage
\begin{figure}[h!]
\centering
\includegraphics[keepaspectratio=true,width=14cm]{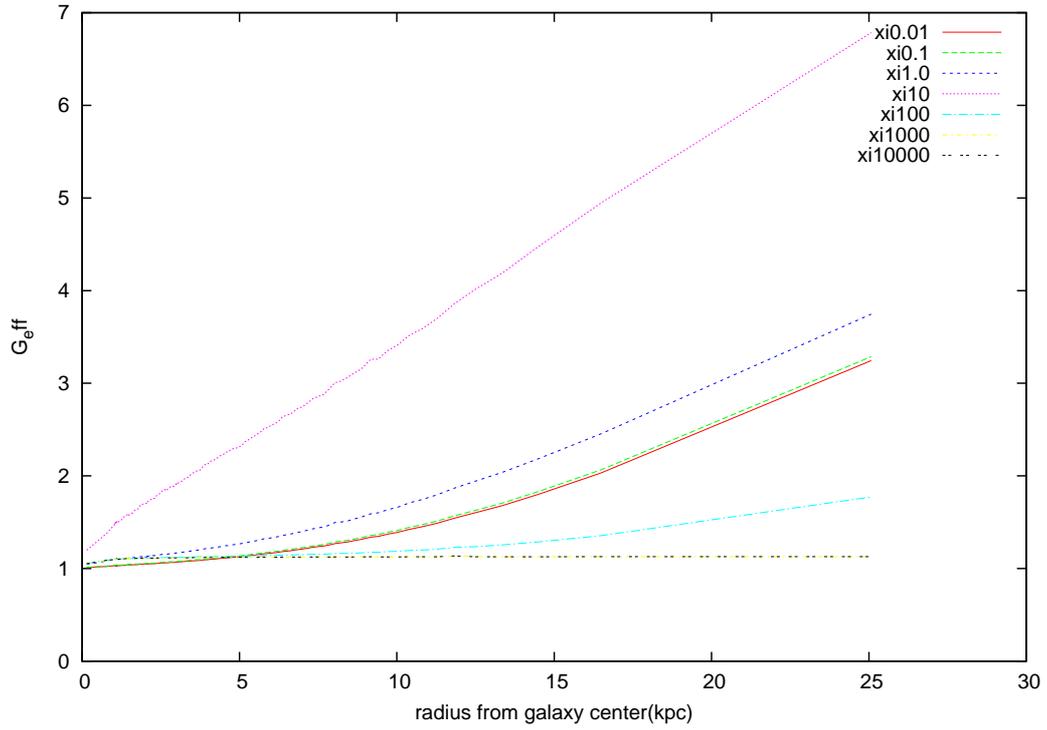}
\caption{The ratio of central acceleration by MOG and the central acceleration by the Newton gravity.We denote this as effective gravitational constant.}
\label{fig:G(r)}
\end{figure}
\begin{figure}[h!]
\centering
\includegraphics[keepaspectratio=true,width=14cm]{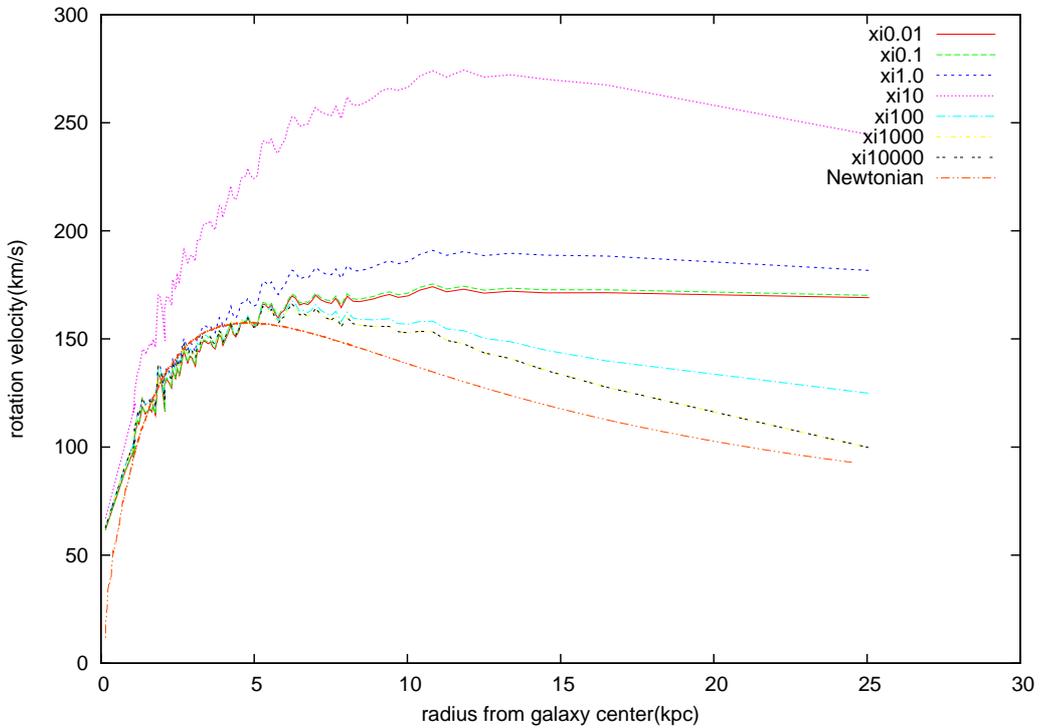}
\caption{The prediction of the galaxy rotation curves under dynamical equilibrium.}
\label{fig:V(r)}
\end{figure}

\newpage

\section{$\xi$ is less than $\mathcal{O}(10^2)$ - Result and consideration}
Then,we calculated the MOG acceleration of each particle in this disk.The result is shown in Fig.\ref{fig:G(r)}.
This figure shows the ratio of central acceleration by the modified gravity and the central acceleration by the Newton gravity as an effective gravitational constant.
$\xi$ in less than $\mathcal{O}(10^2)$,as much as $\xi$ grows large,the effective gravitational constant in galaxy disk scale becomes strong.
If $\xi$ is larger than $\mathcal{O}(10^2)$,as much as $\xi$ grows large,the effective gravitational constant in galaxy disk scale becomes week.
In the case of $\xi=10^4$, the effective gravitational constant accords with Newton gravity.

Fig.\ref{fig:V(r)} shows the prediction of the galaxy rotation curves under dynamical equilibrium about each model.
For a comparison, we calculated the prediction of the galaxy rotation curves on Newton gravity about the same disk.
The result whose $\xi$ is larger than $\mathcal{O}(10^2)$ make a little difference than Newton prediction.

Considering the flat rotation curve observed,$\xi$ should be less than $\mathcal{O}(10^2)$.

%%%%%%%%%%%%%%%%%%%%%%%%%%%%%%%%%%%%%%%%%
\section{Future work}
%%%%%%%%%%%%%%%%%%%%%%%%%%%%%%%%%%%%%%%%%
Now,we are carrying out a large-scale calculation about galactic dynamics and structure formation of universe.
From a dynamical viewpoint, we may obtain a further limitation of $\xi$.
Nurmerical calculation are carried out on Cray XT4 at Center for Computational Astrophysics(CfCA) of National Astronomical Observatory of Japan and SR16000 at YITP in Kyoto University.

%%%   References


\begin{thebibliography}{99}
%
\bibitem[Moffat(2006)]{Moffat} J. W. Moffat,\href{http://iopscience.iop.org/1475-7516/2006/03/004/}{JCAP(03):004, 2006}.

\bibitem[Moffat(2009a]{Toth} J. W. Moffat and V. T. Toth,\href{http://iopscience.iop.org/0264-9381/26/8/085002/}{Class. Quant. Grav. {\bf 26} 085002, 2009}.

\bibitem[Toth(2010)]{Toth2} V. T. Toth,[\href{http://arxiv.org/abs/1011.5174}{[gr-qc]arXiv:1011.5174}].

\bibitem[J. W. Moffat, V. T. Toth(2007)]{MofTot2007}J. W. Moffat and V. T. Toth[\href{http://arxiv.org/abs/0710.0364}{[astro-ph]arXiv:0710.0364}].

\bibitem[Brownstein(2006a)]{Brownstein} J. R. Brownstein and J. W. Moffat, \href{http://iopscience.iop.org/0004-637X/636/2/721/}{Astrophys. J., {\bf 636}, P721-P741, 2006}.

\bibitem[Brownstein(2009)]{Brownstein2} J. R. Brownstein, [\href{http://arxiv.org/abs/0908.0040}{arXiv:0908.0040 [astro-ph]}].

\bibitem[Araujo(2010)]{Araujo} C. S. S. Brandao and J. C. N. de Araujo, Astrophs. J. {\bf 717}, 849 (2010), [\href{http://arxiv.org/abs/1006.1000}{arXiv:1006.1000}].

\bibitem[Brownstein(2006b)]{Brownstein3} J. R. Brownstein and J. W. Moffat, \href{http://onlinelibrary.wiley.com/doi/10.1111/j.1365-2966.2006.09996.x/abstract;jsessionid=519D4F6A7513A302D51EB1FBA6B37530.d01t02}{Mon. Not. R. Astron. Soc., {\bf 367}, 527, 2006}.

\bibitem[Brownstein(2007)]{Brownstein4} J. R. Brownstein and J. W. Moffat, \href{http://onlinelibrary.wiley.com/doi/10.1111/j.1365-2966.2007.12275.x/abstract;jsessionid=A75DD357A96933B056B1B3B37C779685.d03t01}{Mon. Not. R. Astron. Soc., {\bf 382}, 29, 2007}.

\bibitem[Moffat(2010a)]{Moffat2} J. W. Moffat and V. T. Toth, [\href{http://arxiv.org/abs/1005.2685}{arXiv:1005.2685 [astro-ph]}].

\bibitem[Moffat(2008)]{Toth3} J.W. Moffat and V. T. Toth, \href{http://iopscience.iop.org/0004-637X/680/2/1158/}{Astrophys. J., {\bf 680}, 1158, 2008}.

\bibitem[Moffat(2009b)]{Toth4} J. W. Moffat and V. T. Toth, [\href{http://arxiv.org/abs/0901.1927}{arXiv:0901.1927 [astro-ph]}].

\bibitem[Moffat(2009c)]{Toth5} J. W. Moffat and V. T. Toth, \href{http://onlinelibrary.wiley.com/doi/10.1111/j.1365-2966.2009.14876.x/abstract}{Mon. Not. Roy. Astron. Soc., {\bf 397},P.1885-P.1892, 2009}.

\bibitem[Moffat(2010b)]{MoffToth} J. W. Moffat and V. T. Toth, [\href{http://arxiv.org/abs/1001.1564}{arXiv:1001.1564 [gr-qc]}].

\bibitem[Moffat(2009d)]{Toth6} J. W. Moffat and V. T. Toth, Mon. Not. Roy. Astron. Soc., {\bf 395}, L25-L28, 2009.

\bibitem[Takayuki Suzuki(2011)]{my paper1} Takayuki Suzuki, [\href{http://arxiv.org/abs/1110.5420}{arXiv:1110.5420 [astro-ph]}].

\end{thebibliography}
\end{document}